\documentclass[a4paper,12pt]{article}

\usepackage{amsmath}
\usepackage[utf8]{inputenc}
\usepackage{amssymb}
\usepackage{authblk}
\usepackage{graphicx,amssymb,amsfonts}
\usepackage[sort&compress,numbers]{natbib}

\newcommand\be{\begin{equation}}
\newcommand\ee{\end{equation}}
\newcommand\bea{\begin{eqnarray}}
\newcommand\eea{\end{eqnarray}}
\newcommand\del{\partial}

\title{Lorentzian Taub-NUT spacetimes: \\
Misner string charges and the first law}
\author[a,b]{Adel Awad \thanks{a.awad@sci.asu.edu.eg}}
\author[c]{Somaya Eissa\thanks{somaia@sci.cu.edu.eg}}

\affil[a]{\small \it Department of Physics,
Faculty of Science, Ain Shams University, Cairo 11566, Egypt}
\affil[b]{\small \it Centre for Theoretical Physics, the British University in Egypt, El Sherouk City 11837, Egypt}
\affil[c]{\small \it Department of Physics, Faculty of Science,
Cairo University, Giza 12613, Egypt}
\date{}
\date{}
\begin{document}
\maketitle
\begin{abstract}

Motivated by recent activities in Lorentzian Taub-NUT space thermodynamics, we calculate conserved charges of these spacetimes. We find additional mass, nut, angular momentum, electric and magnetic charge densities distributed along Misner string. These additional charges are needed to account for the difference between the values of the above charges at horizon and at infinity. We propose an unconstrained thermodynamical treatment for Taub-NUT spaces, where we introduce the nut charge $n$ as a relevant thermodynamic quantity with its chemical potential $\phi_n$. The internal energy in this treatment is $M-n\phi_n$ rather than the mass $M$. This approach leads to an entropy which is a quarter of the area of the horizon and all thermodynamic quantities satisfy the first law, Gibbs-Duhem relation as well as Smarr's relation. We found a general form of the first law where the quantities depend on an arbitrary parameter. Demanding that the first law is independent of this arbitrary parameter or invariant under electric-magnetic duality leads to a unique form which depends on Misner string electric and magnetic charges. Misner string charges play an essential role in the first law, without them the first law is not satisfied.

\end{abstract}

\section{Introduction}
\label{sec1}
The Taub-NUT spacetime \cite{Taub,NUT} is considered to be one of the most interesting solutions of general relativity. It is known to be an axially symmetric vacuum solution of Einstein's field equations with two parameters, the nut charge and the mass parameter. This spacetime has two Killing vectors, $\del_t$ and $\del_\phi$, the first is time-like, while the second is space-like. For Euclidian metric, the nut charge, $n$ parameterizes a circle bundle over $S^2$, therefore, the spacetime has $\mathbb{R} \times S^3$ topology. The boundary of such a spacetime is topologically non-trivial since it has a non-vanishing first Chern number  which is related to $n$. Because of the last fact, these spacetimes are called asymptotically locally flat rather than asymptotically flat. For the Euclidean metric, this family of solutions contains two classes, the first is called a "nut" solution which has a zero-dimensional fixed-point set of the $U(1)$ isometry generated by $\del_t$. The second is called a bolt solution which has a two-dimensional fixed-point set of the same $U(1)$. For the Lorentzian metric we have only bolt solutions. The Lorentzian and Euclidean Taub-NUT spacetimes have a conical singularity which form a one-dimensional string analogous to Dirac string. It is called Misner string\cite{Misner}. The nut charge is considered to be a magnetic-type mass, in contrast to, the electric-type mass $m$. Therefore, this solution is considered to be a gravitational dyon and the Misner string is a gravitational analogue of Dirac string. Higher dimensional versions of these solutions were introduced in \cite{bais} through constructing a $S^1$ bundle over higher dimensional K${\ddot a}$hler manifolds. Extending these solution in Anti-de-Sitter spaces were introduced in \cite{page+hunter+hawking} in the context of AdS/CFT correspondence as well as its higher dimensional generalizations in \cite{page+pope, adel+andrew} and \cite{adel, mann+stelea}.

There are two important approaches to study this solution, the first is Minser's \cite{Misner}, where one can remove the conical singularity and render the Misner string invisible upon using a large coordinate transformation. This leads to identifying the Euclidian-time direction with periodicity $\beta=8\pi n$, see for example \cite{Ortin-book} and references therein. Considering the thermodynamics in this case, one finds that the horizon radius $r_0$ and the parameter $n$ can not vary independently since $\beta(r_0)=8\pi n$ \cite{clifford98,mann06,clifford14-1,clifford14-2}. As a result, the parameter $n$ can not have its own work-like term in the first law, in contrast with, Kerr or Reissner cases. In addition, the entropy is not proportional to the horizon area. Furthermore, this identification leads to closed time-like curves and causes obstructions to maximal extension of the spacetime, see for example \cite{Kruskal} and references therein. The second approach is due to Bonner \cite{Bonner}, where he keeps Misner string observable, i.e., keeping the conical singularity visible. Bonner keeps the string singularity since he interprets it as a source of angular momentum. Some authors realized the importance of the two approaches as they consider them describing different physical situations, see for example \cite{Dowker}. Several years ago, the authors in \cite{Kruskal} presented some interesting result which showed that Taub-NUT spacetimes can be maximally extended if we abandoned the above periodicity condition. Other authors \cite{Clement1,Clement2} were able to show that the spacetime can be made geodesically complete, without causal pathologies upon abandoning the periodicity condition.

Motivated by the above developments a number of authors revisited Taub-NUT thermodynamics \cite{Durka,mann1,ww,adel_top} trying to formulate it in a manner similar to that of Kerr or Reissner cases, i.e., to have a consistent unconstraint thermodynamics where $\beta \neq 8\pi n$. Let us present briefly some of these approaches that have been introduced by the above authors to construct unconstraint thermodynamics of Taub-NUT spaces, more specifically, a first law with $\beta \neq 8\pi n$. Some of these approaches are: $i)$ Approaches assumes entropy is related to horizon area, or $S= {A_h/ 4}$, in addition, we have extra charges and their chemical potentials. An example of these approaches is the one introduced in \cite{mann1}, where the authors studied the thermodynamics of unconstraint Lorentzian Taub-NUT-AdS. They introduced a new charge ${\cal N}$ related to the NUT charge (it vanishes upon sending $n$ to zero) together with its conjugate chemical potential $\psi$. They were able to write the first law in the form \be dM=TdS+\psi\,d{\cal N}.\ee In their analysis the entropy is a quarter of the horizon area. Further interpretation of this work presented in \cite{Geom_inter.}, where $\psi$ is shown to be proportional to Misner string temperature, and ${\cal N}$ can be interpreted as its entropy. An issue that might be raised here is that since we deal with multi-temperatures system with horizon and Misner temperatures assuming thermal equilibrium might reduce the cohomogeneity of the first law again, through identifying the two temperature, therefore, obtaining the above periodicity condition. Many relevant works followed where authors studied different aspects of this proposal, for example studying the dyonic Taub-NUT-AdS, and its Smarr's relation was covered in \cite{Nutty_dyons,smarr,pando,nutty_dyons2}, while studying Taub-NUT-Kerr and Taub-NUT-Kerr-Newmann was covered in \cite{nutty_rot,FL_rot._taub_nut,rot_nuts}.

Another approach was introduced in \cite{Durka}, where the author calculated various thermodynamical quantities. Among other issues the author stressed on the role of Misner string angular momentum in Taub-NUT-AdS thermodynamics where the first law takes the form
\be dM= Td\left({A \over 4G}\right)+{1\over n} d \left( n\,M \right) - {1 \over 2n} d \,\Xi,\ee where $M$ is the mass, $n$ is the nut charge, $r_{+}$ is the horizon radius and $\Xi={n\,r_{+} \over G} (1+{r_{+}^2 / l^2}+3{n^2 / l^2})$. Although the first law was complete, the last term needed some interpretation/explaination in term of the Taub-NUT thermodynamics quantities as was mentioned at the end of the article. A very similar approach to that in \cite{Durka} was presented in \cite{ww}, where the authors wrote the first law for charged and Kerr-NUT spaces as well. $ii)$ A different approach in which authors considered an entropy which is not related to horizon area, or $S \neq {A_h / 4}$. One of these was introduced in \cite{Geom_inter.,nutty_dyons2} where they consider an entropy on the form \be S= {A \over 4} + {\psi_N\,N_N+\psi_S\,N_S \over T},\ee
where $N_N$ and $N_S$ are the nut charges defined in this approach for the north string and the south string. Also, $\psi_N$, $\psi_S$ are their conjugate chemical potentials which is interpreted as surface gravities along the north and south strings. Another approach introduced in \cite{rot_tn} where they calculate entropy for the Kerr-NUT space using surface charge method \cite{surf-ch} and found that $S \neq {A_h\over 4}$. The mass and angular momentum found in their calculations can be expressed as $M=\alpha n$ and $J= \alpha a n$, where $\alpha$ is some parameter which is not fixed by this method but can be fixed by other means. The first law takes the simple form \be dM = TdS+\Omega dJ.\ee As one can notice here, there is no need to introduce an additional nut charge to the first law, which is the case when the entropy is different from the horizon area.

In this work, we study Lorentzian Taub-NUT solutions, including the neutral Taub-NUT as well as the dyonic Taub-NUT metric. First, we show the existence of extra charge densities between the horizon and radial infinity, for the mass, the electric and magnetic charges as well as angular momentum and nut charges. Second, motivated by the work in \cite{mann1,hunter} and our previous work \cite{adel_top} we revisit the thermodynamics of the above Taub-NUT solutions and propose an alternative treatment which is characterized by the following features. First, we introduce $n$ as a relevant thermodynamic quantity, since it is known to be a magnetic mass. The role of internal energy in this thermodynamical treatment is played by $M-n\phi_n$, rather than the mass $M$, where $\phi_n$ is the chemical potential of the charge $n$. Following \cite{mann1} several authors presented a first law (which was satisfied) with one of the charges (electric or magnetic) at radial infinity but the other at the horizon \cite{Nutty_dyons,smarr,nutty_dyons2}.

Here we show that there is a  more general form of the first law with thermal quantities that depend on some arbitrary parameter. This general case contains the above mentioned cases in literature. Furthermore, this general form is not invariant under electric-magnetic duality. Requiring the first law to be independent of this arbitrary parameter or to be invariant under electric-magnetic duality leads to a unique form. This form can be written in terms of the horizon charges and the charges on Misner string. Indeed, Misner string charges play an important role in the first law, without them the first law is inconsistent.

This article is organized as follows, in section $2$, we review how boundary conditions implies certain thermodynamic ensembles which will be applied to the nut case afterwards. In section $3$, we calculate various conserved charges for the neutral Taub-NUT space, including the mass, nut charge, and angular momentum, where we argue for the existence of charge distributions on the string. Also, We introduce a conserved charge, i.e., $n$ and its chemical potential, then, discuss the thermodynamics of the Taub-NUT which satisfy the first law and Smarr's relation. In section $4$ we calculate the electric and magnetic charges of the dyonic Taub-NUT solution, where we show that they have contributions along the string as well. Also, we calculate various thermodynamical quantities which satisfy, again, the first law and the Smarr's relation. In section $5$, we discuss our treatments and results for the Taub-NUT solutions and reflects on possible extensions and future directions.

\section{Thermodynamical Ensembles}
There is a strong similarity between the nut charge and the magnetic charge of Reissner-Nordstorm solution which will be discussed in the coming sections. They are both magnetic-type quantities \cite{hunter+hawking}, therefore, it is instructive to review the thermodynamics ensembles of Reissner-Nordstorm solution\cite{hawking+ross} with an eye on applying this to Taub-NUT spaces.

Einstein-Maxwell gravitational action for asymptotically flat spacetime is given by
\begin{equation} \label{action}
    I\,=-\frac{1}{16 \pi}\,\int_{\cal M} d^4 x \, \sqrt{-g}\, (R-F^2) -\frac{1}{8\pi} \int_{\del {\cal M}} d^3x
    \, \sqrt{h}\,K.
\end{equation}
where $F_{\mu\nu}=\del_{[\mu}A_{\nu]}$ is the field
strength of a gauge potential $A_{\mu}$ and the last term is the Gibbons-Hawking
boundary term. Here, $h_{ab}$ is the boundary metric and $K$ is the
trace of the extrinsic curvature $K_{ab}$ on the boundary.

The partition function of a charged black hole can be expressed as a path integral over the spatial components of the metric $g_{ij}$ and the gauge field $A_i$
\be Z= \int {\cal D}[A]\,{\cal D}[g]\, e^{-I(g,A)}. \ee
The path integral is subject to boundary conditions that fix $g_{ij}$ and $A_i$ on the boundary, which  determines the type of thermodynamic ensemble of the solution. For example, in the case of a black hole with only a magnetic charge $Q_m$, fixing the boundary value of $A_i$ fixes the magnetic charge, therefore, the partition function is the canonical ensemble partition function \be Z_{can}(\beta,Q_m)=e^{-I}=e^{-\beta F},\ee where, $I$ is the on-shell gravitational action and $F$ is Helmholtz  free energy.
In the electric black hole case the boundary condition fixes the electric potential, $\phi_e$ (chemical potential), rather than the electric charge $Q_e$. Therefore, the
partition function is the grand canonical one, $Z=Z_{grand}(\beta,\phi_e)=e^{-I}=e^{-\beta \Omega}$, where, $\Omega$ is the grand potential. Another possibility, for the electrically charged case, is to describe the system using a canonical ensemble i.e., fixed charge, upon adding the following surface term to the action
\be \tilde{I}=I-{1 \over 4\pi}\int_{\del{\cal M}} d^3x\sqrt{h}\,
n_{a}\,F^{ab}\,A_a. \label{surf-term}\ee This term provides the action with the needed
Legendre transformation to replace its dependence on $\phi_e$ with a dependence on $Q_e$, therefore, one can relate $\tilde{I}$ to the free energy $F$ \be {\tilde{I} \over \beta}=F=U-TS.\ee
In case of dyonic black holes we have a mixed ensemble (i.e., fixed $\phi_e$ and $Q_m$) with a partition function $Z_{mix}(\beta,Q_m,\phi_e)=e^{-\beta \Omega}$, therefore, \be {I \over \beta}=\Omega=U-TS-Q_e\,\phi_e,\ee where, $\Omega$ is the grand potential and the internal energy $U$ is identified with the gravitational mass of the solution $M$. One can describe the dyonic balk holes using a canonical ensemble as well upon adding the above surface term to the action, therefore, fixing both electric and magnetic charges. In all the above cases, the first law has the form
\begin{equation}
    dM = TdS+\phi_m \, dQ_m+\phi_e \, dQ_e, \label{1st law}
\end{equation}
where $S$ is the entropy and $T$ is the temperature.

\section{Taub-NUT Space}
The metric for the Taub-NUT space with a Lorentzian signature is given by
\begin{equation}
    dS^2=-f(r)\,(\,dt+2n(\cos{\theta}+k)\, d\phi\,)^2\,+\,\frac{dr^2}{f(r)}\,+\,(r^2+n^2)\,(\,d\theta^2+\sin^2{\theta}\,d\phi^2), \label{metric}
\end{equation}
where,
\begin{equation*}
    f(r)\,=\,\frac{r^2-n^2-2\,m\,r}{r^2+n^2}.
\end{equation*}
Here $r$ is a radial coordinate, $\phi \in [0,2\pi]$, $\theta \in [0,\pi]$ are angular coordinates, $m$ is the mass parameter, and $n$ is the nut parameter. $k$ is a parameter that determines the position of Misner string \cite{Misner,Clement1,mann1}. Choosing, $k=0$, leads to a conical singularity along the z-axis, which is the location of Misner string \cite{Misner}. Notice that unlike the Euclidian case, the Lorenztian Taub-NUT solution have only bolt solutions, i.e., no nut solutions, since no zero-dimensional fixed-point set of the $U(1)$ isometry of $\del_t$.

As we pointed out in the introduction, this conical singularity can be removed through imposing the periodicity condition, $\beta=8\pi n$, which leaves the Misner string invisible. This condition leads to a relation between $r_0$, the horizon radius, and $n$ the nut charge. Because of this relation there is no new independent work term due to $n$. As a result, the entropy in this setting is not the area of the horizon and the temperature is fixed by the parameter $n$.

Here we are not going to impose the above periodicity condition, therefore, $n$ is independent of $r_0$. Furthermore, we consider "$n$" as a conserved quantity, which provides its own work term in the first law. $n$ is a conserved charge since the nut charge is the magnetic dual of the mass as we will see below. Also in this case, the entropy is the area of the horizon and by taking $n\rightarrow 0$ the expression $T=f'(r_0)/4\pi$ reproduces Schwarzschild temperature in contrast with previous treatments \cite{clifford98,mann06}.

The mass of the above Taub-NUT solution is calculated using Komar's integral. This leads to \be M= -{1 \over 4 \pi}\int_{S_{\infty}^2}\, ^{*}d\xi= m, \ee
where, $\xi=\del_t$ is a time-like Killing vector. For this solution the mass $M$ and the nut charge $n$ are dual quantities, since $n$ is given by \be n= {1 \over 4\pi} \int_{S_{\infty}^2}\, d\xi. \label{n-charge} \ee Comparing the above integrals to electric and magnetic charges expressions, one can see clearly that the mass $M$ is analogous to an electric charge, while $n$ is analogous to a magnetic one. This is why $n$ is called the magnetic mass. The quantity $n$ is clearly conserved, therefore, one expects that it modifies the first law with its own work-like term e.g., $\phi_n\, dn$, where $\phi_n$ is the chemical potential of $n$. Notice that the charge $n$ does not depend on $r_0$ in contrast with the thermodynamics considered in \cite{mann1}. The potential $\phi_n$ can be calculated in a manner similar to that of the magnetic potential, $\phi_m=-\int dr \,[^{*}F]_{0r}$. It takes the form \be \phi_n=-\int dr \, [^{*}d\xi]_{0r}=-{n \over 2\,r_0}.\ee
We will see how this chemical potential is related to other thermodynamic quantities in the coming discussions, but before we do that we show the existence of charge distributions, i.e.,  mass, nut charge and angular momentum along Misner string in addition to their values at infinity.

\subsection{Mass, Angular Momentum and nut Charge}
In this subsection we argue for the existence of mass, angular momentum and nut charge along the z-axis, i.e., along the Misner string using Komar integrals, where $\xi$ and $\chi$ are time-like and space-like Killing vector, respectively.
\subsubsection{Mass}
Apart from the total mass obtained from the flux crossing the boundary at infinity, $M^{\infty}= m$, there are mass densities along z-axis (remember that $k=0$) \footnote{These masses along the z-axis were also discussed in \cite{Manko}.}. These mass densities have a maximum value close to the horizon, $z_0=r_0$, but dies out as $z\rightarrow \infty$. Before showing the existence of these masses in the volume between the horizon and radial infinity surfaces, it is important to remember the following. In the case of Schwarzschild solution, the mass enclosed by the horizon is the same as that enclosed by the sphere at infinity $S^2_{\infty}$. Therefore, there are no mass distributions in the volume between the two surfaces. We will see that for the case with a nut charge $n$, the situation is different, since the two spheres contain different masses. One can use the following two-form to calculate the mass using Komar's integral for both $S^2_{\infty}$ and $S^2_{h}$
\be *d\xi={2n \over r^2+n^2 }f\,dt\wedge dr-{4n^2\, f  \over r^2+n^2 }\cos{\theta}\,dr\wedge d\phi -f'(r^2+n^2)\sin{\theta}d\theta \wedge d\phi. \ee
Clearly, the mass contained in $S^2_{\infty}$ gives $M^{\infty}=m$, while the mass inside the horizon, or, $S^2_{h}$ is \be M^h={r_0^2+n^2\over 2 r_0}. \ee
Furthermore, one can write the horizon mass as $M^h=M^{\infty}-2n\phi_n$. Therefore, these masses are clearly different and this discrepancy reveals the existence of mass distributions between the two surfaces, which is simply the difference $M^{s}=2n\phi_n$. This result becomes less strange upon noticing that the above two-form, is singular along the z-axis, because of the $[*d\xi]_{r\phi}$ component, except at the horizon and the boundary.

Let us calculate the mass along the positive z-axis through calculating the mass flux crossing a cylinder with radius $r\rightarrow 0$ around the axis and above the black hole horizon. We call this cylinder $T_+$. The mass flux crossing this cylinder is given by \be M^{+}= -{1 \over 4 \pi} \int_{T_+}  {^*d\xi}= -{n^2 \over 2r_0}=n\,\phi_n.\ee For the mass on the other side of the z-axis we have \be M^{-}= n\,\phi_n.\ee

This explains why the mass at the horizon is different from that at infinity. The difference is due to the existence of mass distributions inside the volume between the two surfaces and distributed along the z-axis! As one might notice there is a connection between these mass distributions and the conical singularity. We should keep in mind that $k$ is a physical parameter and by changing it we consider different physical systems, since some physical measurable quantities, such as angular momentum depends on it, $J=3knm$. It is known that one can change $k$ through a large coordinate transformation, which relates physically inequivalent cases. In addition, the boundary metric depends on $k$, therefore, one can see that different values of $k$ corresponds to different boundary conditions. Even if we change the position of the string and the mass distributions through changing $k$, one can not set these masses to zero. Unlike the case of Dirac string, where one can use gauge transformations to move the string around. Indeed, $k$ is a physical parameter rather than a pure gauge parameter as was stressed in \cite{Clement1}. As we will see below that this is the case for all conserved charges a long the Misner string not only the mass. Indeed, This is an intriguing aspect of the Taub-NUT solutions which is not shared by other gravitational solutions and was not reported before\footnote{Apart from the author in \cite{Manko} who pointed out the existence of finite mass and infinite angular momentum along Misner string.}.

Notice that the above two-form is perfectly regular at these two surfaces, therefore one can trust the above calculation. Moreover, the value of $M^s=2n\phi_n$ is independent of the value of $k$, i.e., independent of the position of the string. In fact, even if we make Misner string invisible through identifying the time direction, this won't change this result. The existence of this mass density between the two spheres is necessary to account for the difference between the horizon mass and the total mass at infinity. We will see shortly that this is also the case for electric and magnetic charges of the Taub-NUT solutions.\\

\subsubsection{Angular Momentum and Nut Charge}
Apart from the mass-distributions mentioned above we calculate angular momentum and nut charge fluxes at different surfaces to be able to have a consistent picture of this spacetime.
The angular momentum flux contained in some surface $\del\Sigma$ is given by
\be J= -{1 \over 8 \pi}\int_{\del\Sigma}\, ^{*}d\chi, \ee
where, $\chi=\partial_{\phi}$ is is a Killing vector. At the angular momentum fluxes crossing $S^2_{\infty}$, and $S^2_{h}$ are
\be J_{\infty}=0, \hspace{0.25 in} J_{h}=0.\ee
The angular momentum fluxes crossing $T_{\pm}$\footnote{$J_{\pm}$ are divergent quantities, to regularize them one might used the $m=0$ Taub-NUT solution as a background space.} are \be J_{+}=n\, m, \hspace{0.25 in} J_{-}=-n\, m.\ee
Therefore, the total angular momentum on Misner string is $J_{MS}=0$, this is in agreement with \cite{Durka}, see also \cite{Manko}. Notice that the angular velocity at the horizon $\Omega_h$ and at infinity $\Omega_{\infty}$ are vanishing for this spacetime, therefore, one expects that the total angular momentum does not play any role in Taub-NUT thermodynamics. In fact, for the $k\neq 0$ case one can find that $J_{\infty}=3kmn$, which means that different values of the parameter $k$ correspond to different physical systems or setups, i.e., $k$ is a physical parameter. For the nut charge, one obtains \be N_{\infty}=n, \hspace{0.25 in} N_{h}=0.\ee
While on $T_{+}$ and $T_{-}$ we have, \be N_{+}=n/2, \hspace{0.25 in} N_{-}=n/2,\ee therefore, the nut charge are basically coming from the charges on Misner string.

\subsection{Thermodynamics of Taub-NUT Space}
In this subsection, we would like to present an unconstraint treatment of the Taub-NUT thermodynamics, where there is no extra identification of the time direction. Calculating the inverse temperature of this solution one gets
\begin{equation}
  \beta = \frac {4\pi}{f^{\prime}(\,r_0\,)} = {4\pi\,r_0}.
\end{equation}
Following Gibbons and Perry\cite{gibbons-perry} calculations for the Taub-NUT on-shell gravitational action (with Minkowski space-time as a reference space), one obtains
\begin{equation}
   I=\beta \, m/2=\pi\,(r_0^2-n^2). \label{I_n}
\end{equation}
Also, this action can be obtained from that of Taub-NUT-AdS upon using the counter-term method, after setting the cosmological constant to zero as emphasized in \cite{surf99}. Given the above discussion of thermodynamic ensembles, the boundary metric is fixed by boundary conditions, which fixes the nut charge. The partition function is given by \be Z_{can}(\beta,n)=e^{-\beta\, F},\ee
which is describing a canonical ensemble. Therefore, the action is related to the free energy, $I/\beta=F(\beta,n)$. The chemical potential can be obtained from the free energy \be \phi_n=\left({\del F\over \del n}\right)_{T}=-{n \over 2\,r_0},\ee as it should be. This leads to the following relations
\be dF= -S\,dT+\phi_n\, dn.\ee
One can show that \be {I \over \beta}=F(T,n)= M-n\,\phi_n-TS.\label{F_n}\ee But since the relation between the free energy $F$ and the internal energy $U$ is given by $F=U-TS$, one can identify the internal energy in this case
\be U=M-n\,\phi_n =2\pi\,r_0.\ee In fact, this agrees with the standard definition of $U$ as \be U= \langle E \rangle =-\del_{\beta}\, ln Z=\del_{\beta}\, I.\ee
The entropy in this case is obtained from the standard definition
\be S=\beta\, \del_{\beta}\, I-I= \pi\,(r_0^2+n^2),\ee which is the area of the horizon. Now one can see that the first law is satisfied.
\be dU= TdS+ \phi_n\,dn. \label{1st-L_n}\ee
One of the important feature that distinguish this analysis from others works e.g.,\cite{mann1,Geom_inter.} is that the internal energy is not the mass of the system, but $U=M-n\phi_n$. Also, the total change in $F$ is given as
\be dF=-SdT+\phi_n\,dn, \ee where,
\be S=-\left({\del F\over \del T}\right)_n, \hspace{0. in} \phi_n=\left({\del F\over \del n}\right)_T.\ee
\begin{equation}\label{S}
    S=\frac{A_H}{4\,G}.
\end{equation}
Here, we would like to explain and motivate the internal energy obtained above. First, notice that equations (\ref{I_n}) and (\ref{F_n}) implies \be {M}=2 (n \phi_n+TS), \label{Sm_n}\ee which is Smarr's like formula apart from one problem. Remember that we get Smarr's formula from assuming that the internal energy is a homogeneous function of the size of the system $r_0$. But since $M$ and $n$ have dimensions of length one can observe that $U=M-n\phi_n$ rather than $M$ is a consistent definition of internal energy since $U=n\phi_n+2TS$ has the correct scaling. Furthermore, the internal energy by definition depends on the entropy and conserved charges, which is clear from equation (\ref{1st-L_n}). More importantly, this quantity is what one gets from applying the statistical mechanical definition of internal energy or \be U= \langle E \rangle =-\del_{\beta}\, ln Z.\ee
The internal energy is not the mass of the system, but $M-n\phi_n$. To understand this result, let us start with the above solution and set $n=0$, the internal energy is $U=M=r_0/2$. The internal energy is the gravitational energy/mass of the system. Adding a nut charge one finds that the gravitational mass decreases to $r_0\,(1-{n^2\over r_0^2})/2$, but the total or internal energy of the system is $U=M-n\phi_n=r_0/2$. A possible interpretation is that the system uses part of its original energy, i.e., $-n\phi_n$, to deform a Schwarzschild solution to becomes a Taub-Bolt solution. The bottom line is that the internal energy of a gravitational solution is the total available energy of the system which might depends on additional parameters that label the boundary metric rather than the gravitational mass of the system. Our last comments here are about possible different forms of the above first law, which can take the form
\be d(M-2n\phi_n)= TdS- n\,d\phi_n,\ee where $M_h=M-2n\phi_n$ is the horizon mass. Another form reads \be dM= TdS+ dM^s -n d\phi_n,\ee where $M^s=2n\phi_n$ is the mass along Misner string. This shows that part of the mass goes to creates the string mass.\\


\section{Dyonic Taub-NUT Spaces}
In this section we are going to calculate all relevant thermodynamical quantities for the dyonic Taub-NUT solution and show that the first law as well as Smarr's relation are satisfied. In this case the field equations take the form
\be G_{ab}=T_{ab},\ee
where,
\be T_{ab}=F_{ac}\,{F_{b}}^{c}+H_{ac}{H_{b}}^{c}.\ee
The the field strength two-form is related to the one-form potential $A$,  $F=dA$, where $H$ is the Hodge dual of $F$, which is related to the one-form potential $B$, $H={^*F}=dB$. Notice that $T_{ab}$ is invariant under Hodge duality which takes, $F\rightarrow H$ and $H\rightarrow -F$. The first law for dyonic balk holes, with vanishing nut charge, is invariant under this duality as is clear from equation (\ref{1st law}). Also, it enables us to consider $A$ as a fundamental field variable (i.e., its variation produces the field equations) in one case where $B$ is not, but in  the other case, $B$ is the fundamental variable and $A$ is not.
\subsection{A Dyonic Solution}
The dyonic Taub-NUT solution has the same form as the metric in equation (\ref{metric})
where,
\be
    f(r)\,=\,\frac{r^2+p^2+q^2-n^2-2\,m\,r}{r^2+n^2}.
\ee
The one-form gauge potential is given by
\be A= \left(\frac{n\,p\,-\,q\,r\,}{r^2\,+\,n^2}+V \right)\,dt\,+\left(\left(\frac{\,2\,n\,q\,r-p\,(r^2-n^2)}{r^2+n^2} \right) \, \cos \theta+C\right) \,d\phi., \ee
where, $V$ and $C$ are integration constants which will be relevant for the thermodynamics of this solution as we will see shortly. The field strength two-form $F$ is given by
\bea F=&&\frac{2\,n\,p\,r+\,q\,(n^2-r^2)}{(r^2\,+\,n^2)^2}\,dt\wedge dr+ \frac{p\,(n^2-r^2)-2\,n\,q\,r}{r^2+n^2}\,\sin \theta\,d\theta \wedge d\phi \nonumber\\ &&+2\,n\,\cos \theta\,\frac{2\,n\,p\,r+\,q\,(n^2-r^2)}{(r^2\,+\,n^2)^2}\,dr\wedge d\phi.\eea
Notice that
\be F_{r\phi}=2\,n\,\cos(\theta)\, F_{tr}.\ee
The magnetic charge in a spatial region $\Sigma$, with a boundary $\del \Sigma$ is given by \be Q_m=-{1 \over 4\,\pi }\int_{\Sigma}\, dF=-{1 \over 4\,\pi }\int_{\del \Sigma}\, F.\ee
therefore, the magnetic flux at any radius $r$ is \be p(r)= -{1 \over 4\,\pi }\int_{S^2_r}\,F= \frac{p\,(r^2-n^2)+2\,n\,q\,r}{r^2+n^2},\ee
which produces a magnetic charge at $r={\infty}$, \be Q_m^{\infty}=p,\ee
and a magnetic charge at the horizon \be Q_m^h=(p+2\,n\phi_e).\ee
We are going to discuss this discrepancy between the magnetic fluxes at infinity and the horizon in the coming sections.
The dual field strength $H= {^*F}$ is given by
\bea H=&&\frac{-2\,n\,q\,r+\,p\,(n^2-r^2)}{(r^2\,+\,n^2)^2}\,dt\wedge dr+ \frac{q\,(r^2-n^2)-2\,n\,p\,r}{r^2+n^2}\,\sin\theta\,d\theta \wedge d\phi \nonumber\\ &&+2\,n\,\cos\theta\,\frac{-2\,n\,q\,r+\, p\,(n^2-r^2)}{(r^2\,+\,n^2)^2}\,dr\wedge d\phi. \label{H}\eea
Notice that
\be H_{r\phi}=2\,n\,\cos\theta\, H_{tr}.\ee
The electric charge in a spacial region $\Sigma$ is given by \be Q_e={1 \over 4\,\pi }\int_{\Sigma}\, dH={1 \over 4\,\pi }\int_{\del \Sigma}\, H.\ee
Also, the electric flux at any radius $r$ is \be q(r)= {1 \over 4\,\pi }\int_{S^2_r}\,H= \frac{q\,(r^2-n^2)-2\,n\,p\,r}{r^2+n^2},\ee
which produces the following electric charge at infinity \be Q_e^{\infty}=q,\ee
but, at the horizon it takes the form \be Q_e^{h}=(q-2\,n\phi_m).\ee
The electric and magnetic potentials are defined as
\be \phi_e=\Phi_e|_{\infty}-\Phi_e|_{h}=V,\hspace{0.6 in}\Phi_e=A_{\mu}\xi^{\mu},\ee
\be \phi_m=\Phi_m|_{\infty}-\Phi_m|_{h}={p+n\,V \over r_0},\hspace{0.6 in}\Phi_m=B_{\mu}\xi^{\mu},\ee where $\xi$ is a time-like Killing vector field and the one-form $B$ is the solution of $dB=H$, which is  given by
\be B= \left(-\frac{n\,q\,+\,p\,r\,}{r^2\,+\,n^2}+V'\right)\,dt\,+\left(\frac{2\,n\,p\,r+q\,(r^2-n^2)}{r^2+n^2} + C'\right) \, \cos\theta\,d\phi , \ee
where $V'$ and $C'$ are integration constants. We will see later the importance of these integration constants in thermodynamics.

It is known that the thermodynamics of any gravitational solution is a result of Euclidean path integral treatment that imposes certain regularity conditions on the gauge potential $A_{\mu}$. These conditions are obtained from demanding the regularity of the norm of the gauge field, $A^2=A_{\mu}A^{\mu}$, at the horizon and along the z-axis. This quantity can be written as,
\be A^2= { (p+2nV)^2(\cos\theta+s)^2 \over (r^2+n^2)\sin^2\theta}- {(q r-V(r^2+n^2)-np)^2 \over f\,(r^2+n^2)^2},\ee
where we rewrote $C$ in terms of another constant $s$, as $C=s(p+2nV)$. To have a nonsingular potential $A$ on the horizon, the charges $q$, $p$ and the potential $V$ should be related as follows,
\begin{equation}
    q\,=\,\frac{n\,p\,+\,V\,(n^2\,+\,r_0^2)}{r_0}.
\end{equation}
Furthermore, to have a nonsingular potential along the x-axis, or at $\theta=0$ and $\theta=\pi$, we should have two patches for $A$, one is smooth in the northern hemisphere and the other is smooth in the southern hemisphere as in Dirac's monopole case with $s=\mp 1$, or
\be C_{\pm}=\mp(p+2nV),\ee
or \be A_{\pm}^{\phi}={ (p+2nV)(\cos\theta\mp 1) \over (r^2+n^2)\sin^2\theta}.\ee
Notice that the first condition is important for satisfying the first law\footnote{This also was shown in \cite{pando} in a special case where $V=0$. } and the second is needed for obtaining the magnetic charge which is consistent with the above path-integral conditions. This is the magnetic charge in the first law which is different from the total magnetic charge of the system $p$. Regularity of the gauge potential along the z-axis is equivalent to removing the whole z-axis (which carries a magnetic charge $-2n\phi_e$), as a result we get $Q_m=Q_m^h$. This is the magnetic charge in the first law.
Now let us calculate this magnetic charge from the nonsingular one-form $A$ after using Stock's theorem, it gives \be Q_m=- {1 \over 4\pi} \oint A = - {1 \over 4\pi} \left(\int_{north-cap} A_{+} +\int_{south-cap} A_{-}\right)=p+2n\phi_e. \ee
Notice also, by removing the z-axis, the magnetic charge becomes
\be Q_m=-{1 \over 4\,\pi }\int_{\del \Sigma}\, F=-{1 \over 4\,\pi }\left(\int_{S_{\infty}^2}\, F+\int_{T_{+}}\, F +\int_{T_{-}}\, F \right)= p+2n\phi_e.\ee
Our conclusion is that the existence of the nut charge causes a difference between the total magnetic charge and the charge that appears in thermodynamics. The charge in thermodynamics is the result of regularizing the gauge potential $A$ which is required by the Euclidian path integral treatment.



\subsection{Charges on Misner String}

Another intriguing aspect of the Taub-NUT solutions which was not investigated before, is the existence of electric and magnetic charges along the z-axis between the horizon and radial infinity. Indeed, apart from the total charges at infinity, $Q_e^{\infty}= q$, and $Q_m^{\infty}= p$, there are electric and magnetic charge densities along z-axis ($k=0$). These density distributions, again, have maximum values close to the horizon, $z_0=r_0$, but vanishes off as $z\rightarrow \infty$. To show the existence of these charges one can calculate the charges trapped between the horizon and radial infinity. The dual two-form $H$ in equation (\ref{H}) can be used to calculate the electric flux crossing these spheres.
Calculating the electric charge contained in $S^2_{\infty}$ one gets $Q_e^{\infty}=q$, while at the horizon \be Q_e^h=q-2n{p+n\phi_e \over r_0}=q-2n\phi_m. \ee There must be electric charges between the two surfaces, which is equal to $Q_e^{s}=2n\phi_m$. Again, this result is not surprising if we notice that $H$ is singular along the z-axis except at the horizon and the boundary.

On the other hand, one can calculate the electric charge along the positive z-axis through calculating the flux crossing a cylinder with radius $r\rightarrow 0$, around the axis and above the black hole horizon similar to what we did in the mass case. The electric flux crossing this cylinder is given by \be Q_e^{+}= - \int_{T_+}  {H}= n\,\phi_m.\ee For the other side of the z-axis we have \be Q_e^{-}= n\,\phi_m.\ee

Since $H$ is perfectly regular at the horizon and at infinity, therefore one can trust the above calculation. Moreover, even if $k\neq0$ one can check that the charge between the two spheres, i.e., $Q_e^{s}=2n\phi_m$ is independent of the value of $k$, i.e., independent of the position of the string. Similar argument and calculation can be done for the magnetic charge case, where, the magnetic charge contained in $S^2_{\infty}$ is $Q_m^{\infty}=p$, but that at the horizon gives  \be Q_e^h=p+2n\phi_e. \ee Also, \be Q_m^{+}=  \int_{T_+}  {F}= -n\,\phi_e.\ee For the other side of the z-axis we have \be Q_m^{-}= -n\,\phi_e.\ee Similar to the mass case, we emphasize that the existence of these extra charges between the two spheres is necessary to account for the difference in charges collected by the two sphere. This difference is not affected by moving or removing Misner string.

\subsection{Thermodynamics of Dyonic Taub-Nut Solution }
Here we calculate various thermodynamical quantizes for the Taub-NUT Dyonic solution. Its temperature is given by,
\begin{equation} \label{Tch}
\begin{split}
   T = \frac{f^{\prime} (r_0)}{4\pi} = \frac{ (1-V^2)\,r_0^2-\,(p+nV)^2}{4\pi r_0^3}
   \end{split}
\end{equation}

To calculate the on-shell action we follow Gibbons and Perry\cite{gibbons-perry} as in the neutral Taub-NUT case. One gets,
\begin{equation}
    I = \beta\,\frac{(\,m\,r_0^2+(\,p+nV\,)^2-V^2r_0^2)}{2\,r_0}= \beta \Omega.\end{equation}
where, $\Omega$ is the grand potential of the system, which can be written as,
\begin{equation}
    \Omega = U -TS-\Phi_e Q_e = M-n\,\phi_n - TS -\phi_e Q_e. \label{GD}
\end{equation}
Following our discussion on section (2) one can observe that the grand  potential is considered to be $\Omega=\Omega(\beta,n,\phi_e,Q_m)$.
The entropy $S$ is given by
\be S= -\left({\del \Omega \over \del T}\right)_{n,\phi_e,Q_m}=\pi\,(r_0^2+n^2),\ee
which is a quarter of the horizon area.
One can check that the total charges $Q_e=q$ and $Q_m=p$ with the other quantities do not satisfy the first law!\\
\subsubsection{Different forms of the first law and E-M duality}
In this subsection we are going to present first, two different cases which are related by Electric-Magnetic (E-M) duality transformation. These forms of the first law are similar to the cases discussed in \cite{Nutty_dyons,smarr}, where one of the charges is at the horizon and the other at infinity. Then, we are going to present an E-M invariant form of the first law which can be written in terms of horizon charges as well as the charges on Misner string. For all cases the first law, Smarr's relation and Gibbs-Duhem relation (i.e., equation (\ref{GD})) are satisfied.\\
{\bf Case i:}\\
In this case, the first law has a magnetic charge $Q_m=Q_m^{h}=p+2n\phi_e$, which coincide with the charge at the horizon. This leads to a potential \be \phi_m=\left({\del \Omega \over \del Q_m^h}\right)_{T,n,\phi_e}={Q_m^h-n\,\phi_e \over r_0}.\ee
The chemical potential of $n$ is
\be {\phi_n}^{(1)}=\left({\del \Omega \over \del n}\right)_{T,\phi_e,Q_m^h}={1 \over 2 \, r_0}\left[n(\phi_e^2+\phi_m^2-1)-2r_0\phi_e\phi_m\right].\ee
The electric charge that enters the first law is given by
\be Q_e=\left({\del \Omega \over \del \phi_e}\right)_{T,n,Q_m^h}=q.\ee This is the electric charge at radial infinity! These quantities satisfy equation (\ref{GD}) which suggests that the internal energy of the system is \be U=M-n{\phi_n}^{(1)}.\ee This agrees with the standard definition of $U$ in grand canonical or mixed ensemble \be U= \langle E \rangle =\del_{\beta}\, (\beta F).\ee
Now the variation of the grand potential and the first law have the forms
\be d\Omega=-S\,dT+{\phi_n}^{(1)}\,dn+\phi_m\,dQ_m^h-q\,d\phi_e,\ee
\be dU=d(M-n{\phi_n}^{(1)})=T\,dS+{\phi_n}^{(1)}\, dn+\phi_m\,dQ_m^h+\phi_e\, dq.\ee
The electric and magnetic charges appeared in cases $i$ and $ii$ were first derived in \cite{Nutty_dyons}, then in\cite{smarr} but the nut charge, its chemical potential and internal energy were different from ours.
One can check the consistency of the above relations through Smarr's relation which is satisfied and takes the form
\be U=M-n{\phi_n}^{(1)}=2TS+n{\phi_n}^{(1)}+Q_m^h\phi_m+q\phi_e.\ee \\
{\bf Case-ii:}\\
Notice that in case $(i)$, the first law is not invariant under electric-magnetic duality, i.e., under
 \be q\rightarrow p, \hspace{1cm} p \rightarrow -q, \hspace{1cm} \phi_m \rightarrow -\phi_e, \hspace{1cm} \phi_e \rightarrow \phi_m,\ee
as in the $n=0$ case! The first law which is dual to case $(i)$ is given by
\be dU=d(M-n{\phi_n}^{(2)})=T\,dS+{\phi_n}^{(2)}\, dn+\phi_m\,dp+\phi_e\, dQ_e^h,\ee
where, \be {\phi_n}^{(2)}={1 \over 2 \, r_0}\left[n(\phi_e^2+\phi_m^2-1)+2r_0\phi_e\phi_m\right].\ee
Notice that the Hodge duality transforms,  $\phi_e \, dq \rightarrow \phi_m \, dp$,  $\phi_m \, d(p+2n\phi_e) \rightarrow \phi_e \, d(q-2n\phi_m)$ and ${\phi_n}^{(1)} \rightarrow {\phi_n}^{(2)}$. This is another consistent first law which has a magnetic charge $Q_m=p$ but an electric charge $Q_e^h=q-2n\phi_m$. The magnetic potential in this case is
\be \phi_m=\left({\del \Omega \over \del p}\right)_{T,n,\phi_e}={p+n\,\phi_e \over r_0}.\ee
The chemical potential of $n$ is
\be {\phi_n}^{(2)}=\left({\del \Omega \over \del n}\right)_{T,\phi_e,p}.\ee
The electric charge in this case is
\be Q_e=\left({\del \Omega \over \del \phi_e}\right)_{T,n,p}=(q-2\,n\phi_m)=Q_e^h,\ee notice that this is the electric charge at the horizon not at radial infinity!
Now the variations of the grand potential and the first law have the forms
\be d\Omega=-S\,dT+{\phi_n}^{(2)}\,dn+\phi_m\,dp-Q_e^h\,d\phi_e,\ee
\be dU=d(M-n{\phi_n}^{(2)})=T\,dS+{\phi_n}^{(2)}\, dn+\phi_m\,dp+\phi_e\, dQ_e^h.\ee
One can check the consistency of the above relations through Smarr's relation which is satisfied and takes the form
\be U=M-n{\phi_n}^{(2)}=2TS+n{\phi_n}^{(2)}+Q_e^h\phi_e+p\phi_m.\ee
Again, these quantities satisfy equation (\ref{GD}) which leads to \be U=M-n{\phi_n}^{(2)}.\ee Also, this agrees with the definition of $U$ in grand canonical as \be U= \langle E \rangle =\del_{\beta}\, (\beta F).\ee

{\bf An E-M invariant case:}\\
From the above two cases one can observe that there is some redundancy, or ambiguity in the first law. In fact, this becomes clear upon showing that the first law is satisfied for a one parameter family of charges and $\phi_n$ as we will see below. Let us parameterize these quantizes with some parameter $\alpha$, starting with a magnetic charge ${Q_m}^{(\alpha)}=p+\alpha n\phi_e$, which has a magnetic potential \be \phi_m=\left({\del \Omega \over \del Q_m}\right)_{T,n,\phi_e}={Q_m+(1-\alpha)n \phi_e \over r_0}.\ee
The chemical potential of $n$ is
\be {\phi_n}^{(\alpha)}=\left({\del \Omega \over \del n}\right)_{T,\phi_e,Q_m}={n \over 2 \, r_0}\left[n(\phi_m^2+\phi_e^2-1)+2(1-\alpha)\phi_e\phi_m r_0\right].\ee
The electric charge in this case is
\be {Q_e}^{(\alpha)}=\left({\del \Omega \over \del \phi_e}\right)_{T,n,Q_m}=q+(\alpha-2) n\phi_m.\ee
The variation of the grand potential and the first law have the forms
\be d\Omega=-S\,dT+\phi_n\,dn+\phi_m\,dQ_m-Q_e\,d\phi_e,\ee
\be dU=d(M-n{\phi_n}^{(\alpha)})=T\,dS+{\phi_n}^{(\alpha)}\, dn+\phi_m\,d{Q_m}^{(\alpha)}+\phi_e\, d{Q_e}^{(\alpha)}.\ee
One can check the consistency of the above relations through Smarr's relation which  takes the form
\be U=M-n{\phi_n}^{(\alpha)}=2TS+n{\phi_n}^{(\alpha)}+{Q_m}^{(\alpha)}\phi_m+{Q_e}^{(\alpha)}\phi_e.\ee
There are two interesting features here, the first is that E-M duality transformation relates the cases with $\alpha$ and $(2-\alpha)$. Here one should remember that the field equations are invariant under Hodge duality, therefore, it is natural to require a first law that carry this property. It is inetersting to notice that a simple calculation can show that this dependence on $\alpha$ is canceled out if we take \be {\phi_n}^{(\alpha)}={\phi_n}+{n}(1-\alpha)\phi_e\phi_m ,\ee
then the first law takes this form
\be dU=d(M-n{\phi_n})=T\,dS+{\phi_n}\, dn+\phi_m\,d(p+n\phi_e)+\phi_e\, d(q-n\phi_m).\ee
At the same time requiring the invariance of the first law under Hodge duality leads to fixing $\alpha$ to be unity, which is the same $\alpha$-independent form found above!

Furthermore, thermodynamical quantities of this unique expression produce an E-M invariant Smarr's relation as well. In other words, one can get a unique form of the first law by either requiring an $\alpha$-independent expression or an E-M invariant expression.
Another interesting fact is that the above first law can be put in the following form
\bea dU=&&T\,dS+{\phi_n}\, dn+\phi_m\,d(p+2n\phi_e)+\phi_e\, d(q-2n\phi_m) \nonumber\\&&+{1 \over 2}\phi_m\,d(-2n\phi_e)+{1 \over 2}\phi_e\, d(2n\phi_m),\eea
or, \be dU=T\,dS+{\phi_n}\, dn+\phi_m\,dQ_m^h+\phi_e\, dQ_e^h+\phi_m^s\,dQ_m^s+\phi_e^s\, dQ_e^s.\ee

This form is informative since it expresses the first law in terms of the horizon charges (black hole charges) and the Misner string charges, $Q_e^s$ and $Q_m^s$ and their potentials $\phi_m^s=\phi_m/2$ and  $\phi_e^s=\phi_e/2$ \footnote{One can obtain $\phi_e/2 dQ_e^s$ as the space average of the work done to move an infinitesimal charge $dQ_e^s$ from infinity to a finite distance r along the z-axis.}.

To understand more this form of the first law, we should recall our discussion at the end of subsection $4.1$ where we stressed on the fact that the magnetic charges that appear in thermodynamics should be coming from regularizing the one-form potential $A$. This leads to a magnetic charge $Q_m^h=p+2n\phi_e$ not $p$. Also, as a result of having Hode dual field equations, one can choose $B$ to be the field variable instead of $A$. But this leads to imposing the path integral boundary conditions on $B$ instead of $A$. In this case, one can choose the constants, $V'$ and $C'$ in $B$ such that the potential is nonsingular at the horizon and along the z-axis. These values are \be V'={pr_0+nq \over r_0^2+n^2}, \hspace {1.5 cm} C_{\pm}'=\mp (q-2n V' ).\ee
Using Stock's theorem, one get the electric charge \be Q_e={1 \over 4\pi} \oint B =q-2n\phi_m,\ee where $V'=\phi_m$.
This is the electric charge that appears in the first law, which is consistent with E-M duality. Notice that this charge can be produced upon removing the z-axis in the two-form integral \be Q_e={1 \over 4\,\pi }\int_{\del \Sigma}\, H={1 \over 4\,\pi }\left(\int_{S_{\infty}^2}\, H+\int_{T_{+}}\, H +\int_{T_{-}}\, H \right)= q-2n\phi_m.\ee
\section{Conclusion}
We have revisited the thermodynamics of Lorentzian Taub-NUT solutions, including the neutral Taub-NUT as and the dyonic Taub-NUT cases. We argue for the existence of mass, the electric and magnetic charges as well as angular momentum and nut charges a long the z-axis or Misner string. Furthermore, we introduced an alternative treatment which adopt the nut charge $n$ as a conserved charge, since it is known to be dual to the mass. In this treatment the internal energy is $M-n\phi_n$, rather than the mass $M$ as one can observe from the standard definition, \be U= \langle E \rangle =\del_{\beta}\, (\beta F).\ee

We show the existence of a more general form of the first law with quantities that depend on an arbitrary parameter $\alpha$, which includes the cases in literature discussed in \cite{Nutty_dyons,smarr}. By requiring that the first law does not dependent of this arbitrary parameter or to be invariant under electric-magnetic duality we are lead to a unique form which contain horizon charges and the charges on Misner string. The dependence of the first law on horizon charges can be explained as follows. Thermodynamics requires the smoothness of the gauge potential on the horizon and along Misner string. This condition with electric-magnetic invariance of the first law leads to horizon magnetic and electric charges in the first law.
It is clear that Misner string charges do play an important role in the first law, without them the first law is inconsistent. It is also clear that the introduction of a nut charge to the solution changes or shifts the electric and magnetic charges as well as creates various charges along Misner string. It would be interesting to try to understand more the role of these charges on Misner string in thermodynamics especially after adding rotation to the solution. Another extension of this work is to study possible phase transitions for these thermodynamic systems which we would like to report on in the near feature.
\section*{acknowledgments}
A. W. would like to thank M. Alfiky, A. Golovnev and E. Lashin for several
interesting discussion which helped us during the writing of this
work.

\end{document}